\documentclass[12pt,preprint]{aastex}
\usepackage{graphics}
\input epsf

\shorttitle{Central Blue Clumps in Elliptical Galaxies}
\shortauthors{Elmegreen, et al.}

\begin{document}

\title{Central Blue Clumps in Elliptical Galaxies of the Hubble Ultra Deep Field}

\author{Debra Meloy Elmegreen \affil{Vassar College,
Dept. of Physics \& Astronomy, Box 745, Poughkeepsie, NY 12604; e-mail:
elmegreen@vassar.edu} }

\author{Bruce G. Elmegreen \affil{IBM Research Division, T.J. Watson
Research Center, P.O. Box 218, Yorktown Heights, NY 10598, USA; e-mail:
bge@watson.ibm.com} }

\author{Thomas E. Ferguson \affil{Vassar College,
Dept. of Physics \& Astronomy, Box 745, Poughkeepsie, NY 12604;
e-mail:thferguson@vassar.edu}}

\begin{abstract}
Elliptical galaxies larger than 10 pixels in the Hubble Ultra Deep
Field (UDF) were surveyed for internal structure; 30 out of 100 in
a sample of 884 morphologically classified galaxies exhibit large
blue clumps near their centers. Unsharp-masked images of the best
cases are presented. The distributions of the clumps on
color-color and color-magnitude diagrams are about the same as the
distributions of isolated objects in the UDF field with the same
size, suggesting a possible accretion origin. In the few cases
where redshifts are published, the clump masses and star-formation
ages were determined from stellar evolution models, as were the
galaxy masses. The clump mass scales with galaxy mass, probably
because of selection effects, and ranges from $10^6$ M$_\odot$ to
$10^8$ M$_\odot$ for galaxies with masses from $10^{9}$ M$_\odot$
to $10^{11}$ M$_\odot$.  The clump star formation age ranges
between $10^7$ yr and $2\times10^8$ yr. With partial evaporation
and core contraction in the intervening years, some of these
clumps could resemble globular clusters today. Stars that
evaporate will contribute to the field population in the
elliptical galaxy.
\end{abstract}


\section{Introduction}

High resolution images of distant galaxies have revealed many
mergers and interactions (e.g., Abraham et al. 1996; Conselice et
al. 2003; Conselice 2004; Papovich et al. 2005), as might be
expected from hierarchical galaxy build-up (see simulations by
Murali et al. 2002). When mergers trigger star formation, the
resulting galaxies can be blue and clumpy. Hubble Space Telescope
(HST) surveys have confirmed there is a blue population of
elliptical galaxies comprising 15\% to 30\% of the total
ellipticals between z=0.1 and 1 (Menanteau et al. 1999; Im et al.
2001; Gebhardt et al. 2003). Color gradients also indicate blue
cores and other inhomogeneities in over 30\% of resolved
ellipticals in the Tadpole field (Menanteau et al. 2004) and in
the Hubble Deep Fields North and South (Menanteau, Abraham, \&
Ellis 2001). Sanchez et al. (2004) found that 25\% of the AGN
elliptical hosts at redshifts between 0.5 and 1.1 have evidence of
central merging. Elliptical galaxies at $z\sim0.75$ have star
formation time scales less than 1 Gy (Cross et al., 2004). These
observations suggest star-formation and galaxy buildup occurs over
a wide range of redshifts (Franceschini et al. 1998; Menanteau,
Jiminez, \& Matteucci 2001).

Nearby elliptical galaxies also suggest a history of mergers and
accretions because of internal sub-structure (Redna et al. 2004)
and counter-rotating central disks (e.g., Jedrzejewski \&
Schechter 1988; Carollo et al. 1997; Mehlert et al. 1998;
Krajnovic \& Jaffe 2004; Morelli et al. 2004). Double nuclei have
been reported in several ellipticals (Lauer et al. 1996). Enhanced
images sometimes reveal inner spirals or arcs (Lauer et al. 1995)
or dust lanes and filaments (Carollo et al. 1997; Elmegreen et al.
2000; Hunt \& Malkan 2004). Faber et al. (1999) found that typical
elliptical nuclei have a broad range of stellar ages, from 2 to 12
Gy. Marcum, Aars, \& Fanelli (2004) examined the colors and
profiles of nearby isolated ellipticals and found that some have
blue central regions from possible merger events 1-2 Gy ago, even
if they show no unusual morphology.

Here we present data on 100 elliptical-like galaxies in the Hubble
Ultra Deep Field (UDF), of which 30 show evidence of clumps near
their centers in unsharp-masked images.

\section{Data and Results}

The UDF images were observed and processed by Beckwith et al.
(2005) and are available on the Space Telescope Science Institute
(STScI) archive. The UDF consists of images in 4 filters: F435W (B
band, hereafter B$_{435}$; 134880 s exposure), F606W (V band,
V$_{606}$; 135320 s), F775W (i band, i$_{775}$; 347110 s), and
F850LP (z band, z$_{850}$; 346620 s). The images are 10500 x 10500
pixels with a scale of 0.03 arcsec per px.

We classified all galaxies larger than 10 pixels by eye on the
i$_{775}$ image, making a catalog of 884 objects (Elmegreen et al.
2005).  Elliptical galaxies were identified by their optical
appearance, contour plots, and radial profiles. Integrated
magnitudes were measured in a rectangular box defined by the
i$_{775}$ isophotal contours 2$\sigma$ above the sky noise for
each filter (corresponding to a surface brightness of 26.0 mag
arcsec$^{-2}$). The integrated i$_{775}$-z$_{850}$ colors of the
elliptical galaxies range from $-0.35$ to 1.1, with an average of
0.2 to 0.3. The integrated i$_{775}$ magnitudes range from 20.3 to
26.5 mag.

Thirty ellipticals (Table 1) have clumpy structure near their
centers. We enhanced these clumps with unsharp-masked techniques
by subtracting a 5-pixel Gaussian smoothed image from the
originals. Figure 1 shows enhanced images of 8 of them. The figure
also shows contour plots on a logarithmic scale, with the lowest
contour drawn 1$\sigma$ above sky noise, corresponding to a
surface brightness of 26.8 mag arcsec$^{-2}$. In each contour
panel, the lower left-hand number represents the horizontal scale
of the image in arcsec, and the lower right-hand number is the
galaxy identification in the UDF catalog by Beckwith et al., as
listed on the STScI website.
\footnote{http://archive.stsci.edu/pub/hlsp/udf/acs-wfc/h\_udf\_wfc\_VI\_I\_cat.txt}

In most galaxies, the clumps are randomly oriented in the galaxy
with respect to the major axis. Some are
very close to the center, and some are partway out in the galaxy.
Sometimes bar-like or disk-like central structures are seen (as in
UDF25 and UDF9962). These tend to be aligned with the galaxy's
major axis. UDF 2162 shows a ring with three
embedded clumps near the nucleus. UDF900 has an inner spiral with
several clumps along the spiral arms (reminiscent of the spiral in
IC 3328 -- Jerjen, Kalnajs, \& Binggeli 2000).

Figure 2 shows how a typical elliptical, UDF900, appears featureless
on a normal grayscale image.
The radial profile from ellipse fits
is either a de Vaucouleurs r$^{1/4}$ law or a double
power (Sersic) law. For the double power law, the inner slope is
$-0.6$ and the outer is steeper, $-2.4$, as in a
``core'' galaxy (Trujillo et al. 2004). The galaxy's position
angle and ellipticity are shown as a function of radius; the spiral
appears as an abrupt change near the center.

The colors and magnitudes of clumps in each galaxy in Table 1 were
measured by photometry on a box encompassing the peaks, typically
3-5 px across. We subtracted the underlying galaxy intensity from
the clump intensity by measuring the featureless regions adjacent
to each clump. This background subtraction typically made the
clumps 0.1 mag bluer in $i_{775}-z_{850}$ and 0.4 mag dimmer in
i$_{775}$. The measuring errors are about 0.1 mag for the UDF
images. Figure 3 (top left) plots B$_{435}$-V$_{606}$ versus
$i_{775}-z_{850}$ for the clumps and nuclei of the galaxies in
Table 1, and for all 100 elliptical galaxies in integrated light.
For each galaxy, the integrated colors are redder than the clumps
by several tenths of a magnitude. In some galaxies the nuclei are
blue in i$_{775}$-z$_{850}$. Figure 3 (bottom left) shows
i$_{775}$ versus $i_{775}-z_{850}$ for the clumps and the
integrated galaxies. The large spread in color for clumps fainter
than i$_{775}\sim 30$ mag is probably the result of low
signal-to-noise. The clumps average $\sim 4$ mag fainter than the
integrated galaxies.

If the clumps in these elliptical galaxies have been accreted, then
the field should contain isolated objects that are similar in
color and magnitude. The right hand side of Figure 3 shows sources
in the UDF catalog that have sizes between 3 and 4 pixels FWHM.
This is approximately the size range for the clumps in our
elliptical galaxy sample. The color-magnitude and color-color
distributions of the tiny field clumps resemble the distributions
for the elliptical galaxy clumps, lending credibility to the
accretion model.

The redshifts of four of our clumpy ellipticals were measured by
Vanzella et al. (2004) and are listed in Table \ref{tab:clumps}.
They range from $z=0.218$ to $1.317$, illustrating how clump
formation or accretion occurs during a wide range of epochs. The
same survey found the redshifts of 2 other ellipticals that do not
have obvious clumps, UDF 3088 and UDF 9264. Using these redshifts,
we converted the apparent magnitudes into absolute i$_{775}$ AB
magnitudes and obtained the major axis galaxy diameters,
considering a standard $\Lambda=0.7$ cosmology. The
background-subtracted B$_{435}$-V$_{606}$, V$_{606}$-i$_{775}$,
and i$_{775}$-z$_{850}$ colors and the i$_{775}$-band magnitudes
of the clumps and whole galaxies were also compared to Bruzual \&
Charlot (2003) evolution models in order to estimate the masses,
ages, and star formation decay times, assuming an exponential star
formation rate. Two values of the internal dust extinction were
taken, one from Rowan-Robinson (2003) for the appropriate
redshifts and another four times larger for comparison. The
Rowan-Robinson extinctions are slightly lower than others measured
for spirals at the same redshift (Adelberger \& Steidel 2000;
Papovich, Dickinson, \& Ferguson 2001), but they may be a good
approximation for ellipticals;  the second case with $4\times$
higher values should bracket most previous results. Absorption
from intervening hydrogen lines and continuum were included, from
Madau (1995). The models are discussed in Elmegreen \& Elmegreen
(2005).

The model results are in Table \ref{tab:clumps}. The clump masses
range from $6\times10^5$ M$_\odot$ to $5\times10^7$ M$_\odot$,
with slightly larger values for higher extinctions (as shown in
parentheses). The star formation ages of the clumps are typically
$\sim10^8$ yr for low extinction and several$\times10^7$ yr for
high extinction. Clump diameters are several hundred pc to a kpc.
Only clumps that gave close fits to the model colors (with an rms
deviation from all 3 colors less than several tenths) are listed
with a mass and age.  For the whole galaxies, the star formation
ages are much longer than for the clumps, but still less than the
age of the Universe at that redshift. This suggests the
ellipticals either have lingering star formation or contain other,
more dissolved, clumps spanning a wide range of ages. To give an
upper limit to the whole galaxy mass, solutions were also found
for assumed ages equal to the galaxy age if star formation started
at $z=6$ and then decayed exponentially. In these cases, the decay
time was chosen to match the colors.  These results are given in
square brackets.  Typically the galaxy mass increases by a factor
of $\sim3$ with this assumption. Considering the large measurement
and model uncertainties, these results should be viewed as order
of magnitude estimates.

The clump ages appear to be
comparable to or less than the orbital accretion times for the
elliptical galaxies, which typically exceed several hundred My.
This implies that most of the young stars formed after the clumps
entered the galaxies, if the accretion model is correct.  We note
from Figure 3 that the elliptical galaxy clumps lie at the lower
envelope of the field clump B$_{435}$-V$_{606}$ color
distribution, consistent with triggered star formation during
accretion.  Alternatively, the clumps could
have formed inside gas clouds that were already in the
ellipiticals.

The clump masses are an order of magnitude larger than globular
cluster masses. Considering the likely evaporation of cluster
stars in the intervening time and the likely central concentration
of young stars inside these clumps, some of these objects could
turn into globular clusters. This situation slightly resembles
models where globular clusters and ellipticals both form during
gas-rich mergers (e.g., Ashman \& Zepf 1992). However, the clumps
here are younger than the relaxation times of the host galaxies,
so if they formed when the ellipticals formed, then star formation
had to continue for a relatively long time. The observations more
resemble models where globular clusters enter a galaxy as parts of
dwarf galaxies (e.g., Searle \& Zinn 1978).  If the observed
clumps evaporate more completely, then the elliptical galaxies
will become more smooth, leaving only faint remnants of accretion
such as those discussed in the introduction.

\acknowledgments

We are grateful to Vassar College for undergraduate research
support for T.E.F. and for support from the Salmon fund for
D.M.E.. B.G.E. is supported by the National Science Foundation
through grant AST-0205097.


\clearpage

\begin{deluxetable}{lcccc}
\tabletypesize{\scriptsize} \tablecaption{Elliptical Galaxies with
Clumps \label{tab:fits}}
\tablewidth{0pt}
\tablehead{
\colhead{UDF no.} &
\colhead{Features} &
\colhead{UDF no.} &
\colhead{Features} \\ }
\startdata
25&central bar; 5 nearby clumps&3174&2 outer clumps\\
100&3 near-nuclear clumps&3677&elongated nucleus\\
153&1 near-nuclear clump, 1 mid-disk clump&4142&elongated nucleus plus 1 near-nuclear clump\\
206&1 near-nuclear clump&4389&central arc plus 2 clumps\\
221&4 near-nuclear clumps&4445&1 nuclear clump, outer streamers\\
703&2 near-nuclear clumps&4551&elongated nucleus\\
900&central spiral with 10 clumps&6018&lumpy nucleus plus 1 nuclear clump\\
901&3 nuclear clumps plus 3 clumps outside nucleus&6027&2 near-nuclear clumps plus 2 outer clump arcs\\
1088&nuclear clump plus 2 outer clumps&6288&2 big nuclear clumps plus 3 smaller clumps\\
1564&1 near-nuclear clump plus 2 outer clumps&8138&2 nuclear clumps\\
1607&elongated lumpy nucleus &8680&elongated nucleus\\
1727&elongated nucleus&9962&nuclear bar; 2 end-of-bar clumps\\
1960&2 near-nuclear arcs&E1\tablenotemark{a}&1 near-nuclear clump\\
2162&nuclear arc with 3 clumps&E2\tablenotemark{a}&1 nuclear clump plus 1 outer arc\\
2974&1 near-nuclear clump&E3\tablenotemark{a}&elongated nucleus plus 1 near-nuclear clump\\
\enddata
\tablenotetext{a}{no UDF numbers are given for three object in our table; we designate them by E1, E2, and E3.
Their coordinates are,
E1: 3$^{\rm h}$32$^{\rm m}$30.4637$^{\rm s}$, -27$^\circ$48$^\prime$3.068$^{\prime\prime}$;
E2: 3$^{\rm h}$32$^{\rm m}$42.5604$^{\rm s}$, -27$^\circ$45$^{\prime}$50.196$^{\prime\prime}$;
E3: 3$^{\rm h}$32$^{\rm m}$42.2572$^{\rm s}$, -27$^\circ$49$^\prime$15.137$^{\prime\prime}$}
\end{deluxetable}

\clearpage

\newpage
\begin{deluxetable}{lccccccccccc}
\tabletypesize{\scriptsize} \tablecaption{Properties for Galaxies
with Spectroscopic Redshifts \label{tab:clumps}} \tablewidth{0pt}
\tablehead{
\colhead{Object} &
\colhead{z\tablenotemark{a}} &
\colhead{i} &
\colhead{B-V}&
\colhead{V-i}&
\colhead{i-z}&
\colhead{I}&
\colhead{Diam.}&
\colhead{Age Fit\tablenotemark{b}}&
\colhead{Age Max\tablenotemark{c}}&
\colhead{log Mass\tablenotemark{b}}&
\colhead{log Mass Max\tablenotemark{c}}\\
\colhead{} & \colhead{} & \colhead{mag}& \colhead{mag}&
\colhead{mag}& \colhead{mag}& \colhead{mag}& \colhead{kpc}&
\colhead{My}& \colhead{My}& \colhead{M$_\odot$}&
\colhead{M$_\odot$}\\} \startdata
UDF 153&0.98&21.59&2.05&1.45&0.88&$-22.5$&29&1900(1900)&5100&11.1(11.8)&11.6(12.2)\\
\hspace{0.1in}Clump 1&&28.53&1.03&0.48&0.08&$-15.5$&0.8&180(40)&&7.3(7.5)&\\
\hspace{0.1in}Clump 2&&26.67&0.80&0.21&$-0.16$&$-17.4$&1.7&90(14)&&7.7(7.8)&\\
UDF 3088&0.127&25.94&1.37&1.02&0.73&$-13.0$&1.6&4800(1300)&11000&7.1(7.1)&7.4(7.4)\\
UDF 4142&0.737&21.70&0.60&0.66&0.20&$-21.6$&15&120(50)&6200&9.5(9.7)&10.3(11.0)\\
\hspace{0.1in}Clump 1&&29.06&$-0.05$&0.10&0.67&$-14.2$&0.8&13(10)&&5.8(6.4)&\\
UDF 6027&1.317&23.80&1.12&1.23&0.91&$-21.1$&16&1000(700)&3900&10.5(11.1)&10.8(11.5)\\
\hspace{0.1in}Clump 1&&30.21&$-0.16$&0.03&$-0.21$&$-14.7$&1.0&20(\nodata)&&6.0(\nodata)&\\
\hspace{0.1in}Clump 2&&30.87&6.26&0.84&$-0.35$&$-14.0$&0.9&\nodata&&\nodata&\\
\hspace{0.1in}Clump 3&&31.06&$-2.54$&3.23&$-0.03$&$-13.8$&1.1&\nodata&&\nodata&\\
\hspace{0.1in}Clump 4&&30.08&$-0.00$&$-0.46$&$-1.77$&$-14.8$&1.4&\nodata&&\nodata&\\
UDF 9264&1.096&21.96&1.94&1.53&1.08&$-22.4$&32&1900(1300)&4600&11.2(11.8)&11.6(12.2)\\
E2&0.218&22.36&0.28&1.04&0.18&$-17.8$&6&1200(45)&10000&8.4(8.3)&8.9(9.4)\\
\hspace{0.1in}Clump 1&&27.75&\nodata&\nodata&$-0.26$&$-12.4$&0.4&\nodata&&\nodata&\\



\enddata
\tablenotetext{a} {Redshifts from Vanzella et al. (2004)}
\tablenotetext{b} {Values in parentheses are for $4\times$ the
extinction compared to Rowan-Robinson (2003).} \tablenotetext{c}
{Values are best fits assuming star formation began at redshift
$z=6$.}
\end{deluxetable}



\clearpage

\begin{figure}\epsscale{0.8}

\plotone{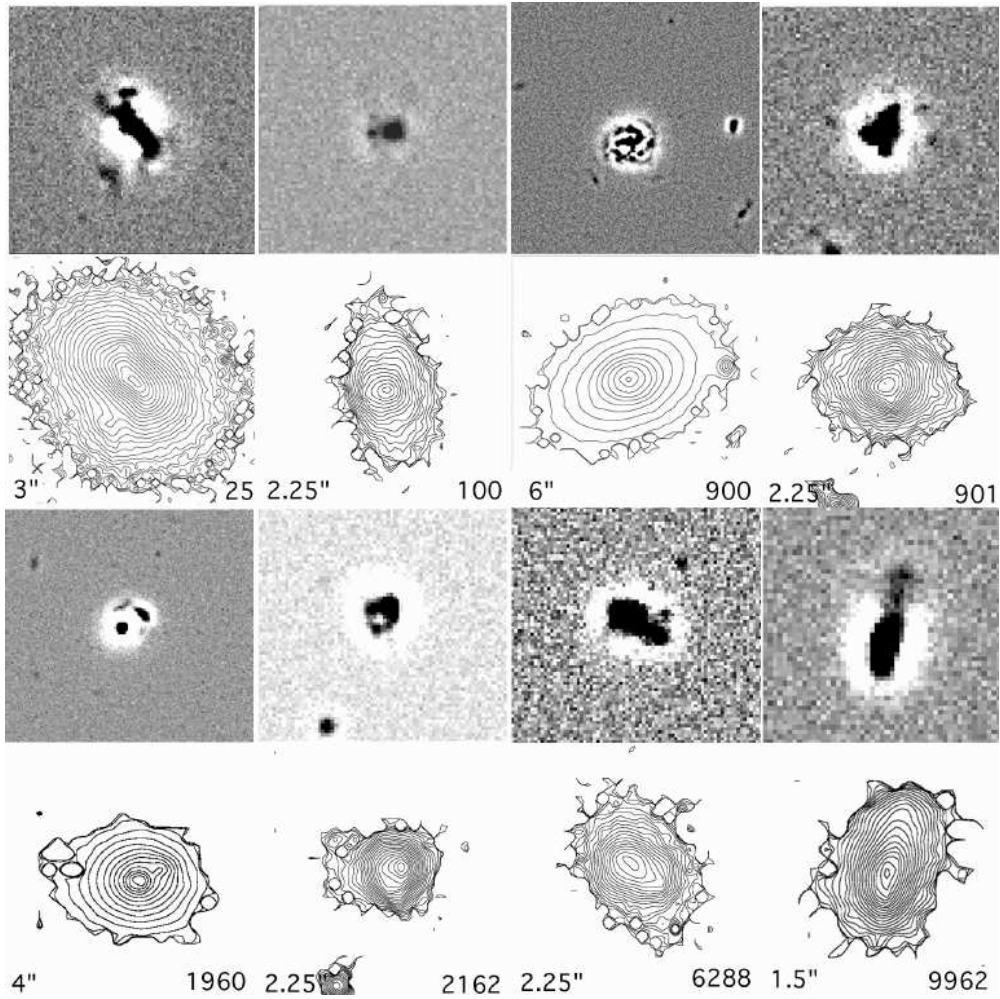} \caption{Unsharp-masked i$_{775}$ images and
logarithmic contour plots are shown for 8 of the 30 galaxies in
Table 1, which have central clumpy features.}\end{figure}

\clearpage

\begin{figure}\epsscale{0.8}
\plotone{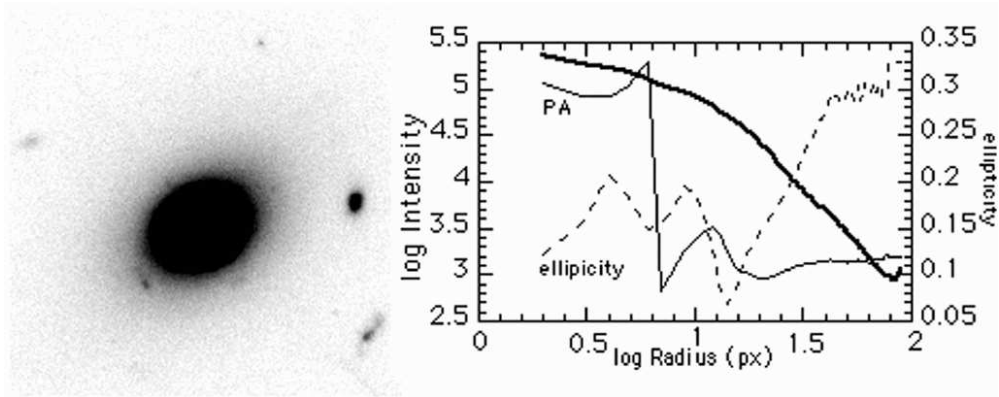} \caption{The i$_{775}$ grayscale image of UDF 900
is shown with the mean intensity, position angle, and
ellipticity as a function of radius, based on ellipse fits. The
spiral structure evident in Figure 1 ends at a radius of about 30
pixels.}\end{figure}

\clearpage

\begin{figure}\epsscale{0.8}
\plotone{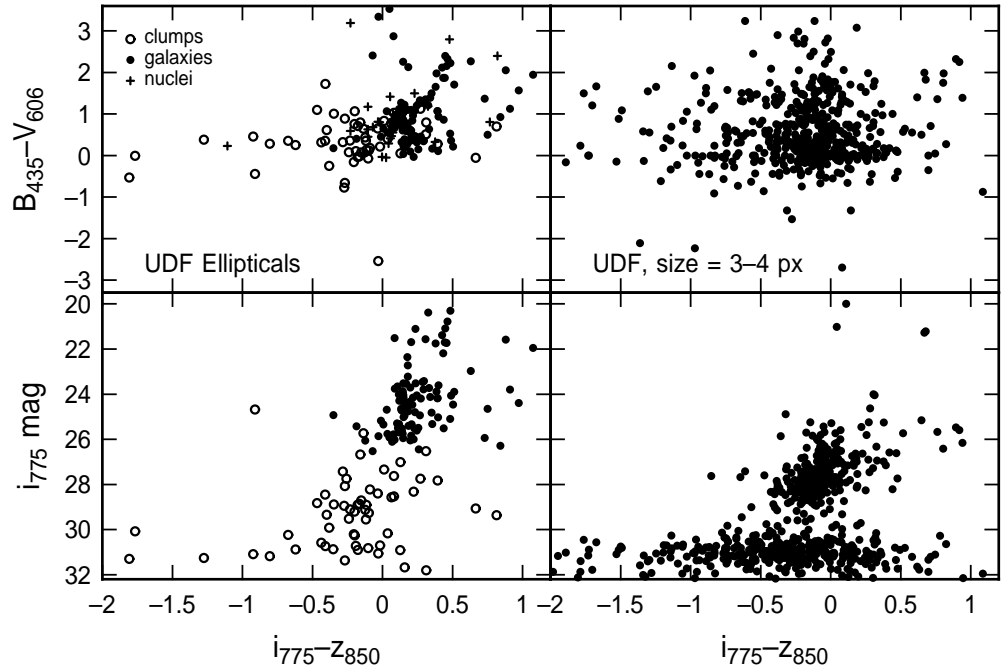} \caption{Top left: Color-color distributions of
the clumps and nuclei for galaxies in Table 1, and for all 100
elliptical galaxies in integrated light. Bottom left:
Color-magnitude distributions of the clumps (open circles) and
integrated elliptical galaxies (dots). Right: Similar
distributions for all UDF-catalogued objects with FWHM between 3
and 4 px.  }
\end{figure}

\end{document}